\documentclass[conference]{IEEEtran}
\IEEEoverridecommandlockouts
\usepackage[compress,sort]{cite}
\usepackage{amsmath}
\usepackage{amssymb,amsfonts}
\usepackage{algorithmic}
\usepackage{graphicx}
\usepackage{textcomp}
\usepackage{float}
\usepackage{pifont}
\usepackage{tabularx}
\usepackage{booktabs}
\usepackage{comment}
\newcommand{\cmark}{\ding{51}}   
\newcommand{\xmark}{\ding{55}}   
\usepackage{graphicx} 
\usepackage{xcolor}
\def\BibTeX{{\rm B\kern-.05em{\sc i\kern-.025em b}\kern-.08em
    T\kern-.1667em\lower.7ex\hbox{E}\kern-.125emX}}

\begin{document}
\pagestyle{plain}   

\title{MuMeNet: A Network Simulator for Musical Metaverse Communications\\}

\author{  
\IEEEauthorblockN{
    Ali Al Housseini\IEEEauthorrefmark{1}\IEEEauthorrefmark{3},
    Jaime Llorca\IEEEauthorrefmark{2}\IEEEauthorrefmark{4},
    Luca Turchet\IEEEauthorrefmark{2},
    Tiziano Leidi\IEEEauthorrefmark{3},
    Cristina Rottondi\IEEEauthorrefmark{1},
    Omran Ayoub\IEEEauthorrefmark{3}\\
    }

  \IEEEauthorblockA{\IEEEauthorrefmark{1}
    Politecnico di Torino, Turin, Italy, \IEEEauthorrefmark{2}
    University of Trento, Trento, Italy}
    \IEEEauthorblockA{\IEEEauthorrefmark{3}
    University of Applied Sciences and Arts of Southern Switzerland, Lugano, Switzerland}
    \IEEEauthorblockA{\IEEEauthorrefmark{4}
    Centre Tecnologic de Telecomunicacions de Catalunya (CTTC/CERCA), Castelldefels, Spain.}

  \small ali.alhousseini@supsi.ch
}

\setlength{\abovecaptionskip}{4pt}
\setlength{\belowcaptionskip}{4pt}
\maketitle

\begin{abstract}
The Metaverse, a shared and spatially organized digital continuum, is transforming various industries, with music emerging as a leading use case. Live concerts, collaborative composition, and interactive experiences are driving the Musical Metaverse (MM), but the requirements of the underlying network and service infrastructures hinder its growth. These challenges underscore the need for a novel modeling and simulation paradigm tailored to the unique characteristics of MM sessions, along with specialized service provisioning strategies capable of capturing their interactive, heterogeneous, and multicast-oriented nature. To this end, we make a first attempt to formally model and analyze the problem of service provisioning for MM sessions in 5G/6G networks. We first formalize service and network graph models for the MM, using \lq \lq live audience interaction in a virtual concert" as a reference scenario. We then present \emph{MuMeNet}, a novel discrete-event network simulator specifically tailored to the requirements and the traffic dynamics of the MM. 
We showcase the effectiveness of \emph{MuMeNet} by running a 
linear programming based orchestration policy on the reference scenario 
and providing performance analysis under realistic MM workloads.
\end{abstract}


\section{Introduction}
\label{sec:intro}
The \emph{Metaverse} refers to a vision of a shared, spatially organized digital continuum in which individuals experience an embodied presence and interaction across interconnected and heterogeneous virtual, augmented and physical environments \cite{metaverse}. Early deployments of the Metaverse have centered on gaming, retail, and digital twins for industrial monitoring, simulation, and optimization, serving as testbeds for immersive interaction and real-time collaboration applications \cite{cai2022compute}. 

More recently, new forms of the Metaverse are emerging, among them, the \emph{Musical Metaverse} (MM), which aims to enable real-time, interactive experiences such as live virtual concerts, collaborative music creation, and shared listening sessions \cite{MMsurvey}.
From a technical perspective, the MM represents an interoperable, persistent network of multi-user environments that merge physical reality with digital reality. It is based upon the convergence of Musical eXtended Reality (XR) \cite{MusicInXR} and Internet of Musical Things (IoMusT) \cite{IoMUST} technologies that enable multisensory, networked musical interactions, not only among musicians, audiences, and performers, but also with virtual environments and intelligent instruments
\cite{MMsurvey}.

Realizing such deeply immersive and time-sensitive experiences places unprecedented demands on the network infrastructure. Unlike conventional media applications, MM sessions involve synchronized delivery of high-fidelity audio and video alongside real-time control signals \cite{RTMusicHptics}, gestural and sensor data \cite{emotions}, and haptic feedback \cite{haptics}. Achieving musical co-presence, i.e., the perception of synchronous musical interplay, requires ultra-low end-to-end latency (in the order of a few milliseconds), near-zero jitter, and bidirectional data exchange \cite{latencyFactors, MusicInXR}. These stringent requirements challenge existing communication frameworks and necessitate a distributed, responsive infrastructure encompassing 5G/6G technologies, edge computing, and adaptive bitrate control \cite{MMsurvey, metaverseComm}. 

To meet Quality of Service (QoS) and Quality of Experience (QoE) expectations, MM sessions must
\emph{adapt} in real time to changing network conditions and explicitly \emph{account} for the concurrent multimodal traffic \cite{MMsurvey}.
These challenges underscore the need for novel modeling and simulation methods tailored to the unique aforementioned characteristics of MM sessions, along with specialized service provisioning and orchestration strategies capable of capturing their real-time, interactive, and multicast-oriented nature \cite{idago}. 

To model, simulate, and analyze the problem of service provisioning in the MM, several aspects should be considered:

\begin{itemize}
    \item \emph{Scale and Heterogeneity}: A realistic MM session spans thousands of geographically dispersed users, multiple edge tiers, backbone transport, and cloud data centers. Recreating that hierarchy in hardware would demand a large number of servers, equipments, and investments.

    \item \emph{Observability and repeatability}: Field measurements can reveal end-to-end delays, but make it difficult to expose what happens \emph{inside} each compute node (e.g., buffers); nor can they be replayed under identical load. A simulation environment, by contrast, allows one to have deterministic control over every link, timestamp, and flow, enabling fair comparisons and statistically significant ablation studies.

    \item \emph{Risk and iteration speed}: Early-stage policies often violate QoS constraints; running them live would produce audible glitches or broken sessions. Simulation environment offers a low-risk sandbox in which unsafe ideas can fail fast and guide the next design cycle.
\end{itemize}

In this work, we make a first attempt to formally model, analyze, and simulate the problem of service provisioning for MM sessions in 5G/6G networks. To this end, we offer the following contributions:

\begin{itemize}
    \item We design a graph-based model for MM services that describes the interaction between MM components, and illustrate it for the specific scenario of live audience interaction in a virtual concert;
    \item We leverage the CNFlow framework to build \emph{service graph} (SG) and \emph{cloud-network graph} (CNG) mathematical models for the MM that reconceptualizes how services are mapped on the network infrastructure;
    \item We introduce \emph{MuMeNet}, a discrete-event network simulator specifically tailored to the requirements and traffic dynamics of the MM that allows embedding (i.e., allocation of resources) dynamic SGs while analyzing the cross-stream synchronization. 
    \item Using the proposed modeling, we formulate the joint placement and routing of MM SGs onto a CNG as a Mixed-Integer Linear Program (MILP) and demonstrate its effectiveness across diverse network instances and traffic scenarios simulated via MuMeNet.
\end{itemize}


The remainder of this paper is organized as follows. Section~\ref{sec:related} discusses related works. Section~\ref{sec:musmopt} models the MM scenario. Section~\ref{sec:simulator} describes the architecture of \textsc{MuMeNet}.   Section~\ref{sec:ilp} presents the MILP. Section~\ref{sec:results} discusses numerical results, and Section~\ref{sec:conclusion} concludes the paper.

\section{Related Work and Technical Gaps}\label{sec:related}

\subsection{Network Communication and Resource Allocation in the Metaverse}

Early visions depict the Metaverse as the Internet's next evolutionary stage, enabling users to work, play, and socialize in persistent, fully immersive 3D spaces \cite{tang}. Delivering such experiences at scale places unprecedented pressure on telecommunication networks as they must support thousands of concurrent users, stream multi-sensory XR content at gigabit rates, and keep motion-to-photon delays below 20 ms to sustain presence \cite{metaverseComm}. Tang \textit{et al.} outline a 6G roadmap that combines intelligent sensing, integrated space–air–ground networking and edge computing to achieve the Ultra-Reliable Low-Latency Communication (URLLC) essential for the Metaverse \cite{tang}. In \cite{qosImpact}, authors conducted empirical studies that confirm that even modest latency spikes or packet loss degrade QoE, consequently producing motion lag and user dizziness in networked Virtual Reality (VR). It is thus envisioned that next-generation infrastructures will be re-engineered around 5G/6G slices and edge intelligence to satisfy the extreme bandwidth, latency, and reliability demands of immersive, real-time Metaverse applications. Complementing this network focus, Van \textit{et al.} propose a digital-twin-enabled architecture that offloads compute-intensive tasks (e.g., physics updates, rendering) to multi-access edge servers and caches popular results, jointly optimizing communication, computation, and storage to trim end-to-end delay \cite{van}. 

Beyond architectural enablers, a significant body of recent research addresses resource allocation algorithms for efficiently managing network and computing resources in Metaverse environments. In \cite{zhao}, the authors introduce a human-centric resource allocation framework for wireless Metaverse applications that explicitly optimizes perceived user utility relative to the cost of resources used. Their system jointly allocates communication bandwidth, computing power (for rendering), and even adjusts the streaming video resolution to maximize a utility-cost ratio, essentially delivering the best experience per unit of cost in energy and latency. 

A recent line of work by Chu. \textit{et al.} tackles generic Metaverse resource management through a two-stage evolution. Their main contribution, \emph{MetaSlicing} \cite{chu2023metaslicing}, decomposes each application into fine-grained functions and clusters applications that share those functions into \emph{MetaInstances}, so that one instantiation can be transparently reused across slices. 

While such contributions are notable, the comprehensive orchestration of resource-intensive Metaverse applications necessitates novel modeling approaches that can accurately capture their intricate processing graph structures, multimodal flow scaling behavior, and efficient sharing and replication of real-time data streams \cite{idago}. For this reason, we leverage the Cloud Network Flow (CNFlow) modeling and optimization framework \cite{CNFLOW,idago,carryinf}, which offers a robust abstraction for modeling and optimizing the end-to-end orchestration of next-generation media services over distributed cloud-integrated networks.

Despite the rapid progress in research on Metaverse communications and service provisioning, most studies to date validate their proposals via analysis or simulation (e.g., custom network simulators or trace-driven experiments). There is a lack of standardized design and modeling of Metaverse slices, as well as of testbeds and emulation platforms for Metaverse networking. This raises questions about how solutions will perform under real network conditions, with unpredictable user behaviors and interference. Developing realistic prototypes and common simulation benchmarks for Metaverse scenarios remains an open challenge.

\subsection{Comparison between Existing Network Simulators}

\paragraph*{Packet-level network simulators} Several works extend classical packet simulators to Metaverse use-cases. Wang~\emph{et al.} \cite{wang2024experimental} connect a cloud renderer to a HoloLens 2 client in a \lq \lq Virtual City" built on \texttt{ns-3} \cite{ns3}, assessing end-to-end delay and throughput for remote AR rendering. Lecci~\emph{et al.} \cite{lecci} contribute an ns-3 XR traffic generator that replays head-movement traces as bursty UDP flows, emulating the high data rate of VR video, while audio support is left as future work. Recent research prototypes even couple network simulators (OMNET++ \cite{omnetpp}) with 3D engines (e.g., Unity) to create holistic simulations aimed at analyzing how dense user motion affects 5G links in a virtual event arena. While these attempts enable a faithful evaluation of congestion control and scheduling, their primary limitation is architectural: these tools assume \emph{static} network topologies and treat packets as homogeneous byte streams, lacking the abstractions necessary to represent SGs or modality-specific latency constraints.

\paragraph*{Cloud/edge simulators}
To explore service placement and resource allocation policies, the research community has widely adopted the CloudSim family of simulators, whose internal models follow the classic Virtual Network Embedding (VNE) abstraction \cite{fischer}. CloudSim Plus \cite{cloudsim} focuses on data-center virtual machine scheduling, while fog/edge variants reuse the same VNE kernel in distributed settings. Although these frameworks excel at graph-layer optimization, they cannot express information-centered replication, mixed-cast flows, or millisecond-level latency variance, capabilities that MM workloads require. In addition, critical network dynamics, such as packet jitter, burst loss, and congestion effects (the primary causes of musical desynchronization), are rendered invisible. As a result, a placement strategy that appears optimal in simulation may prove unusable once the dynamics of a real-world network are considered.


The work in \cite{dual} provides an initial solution to incorporate the CNFlow framework into ns-3 and allow service orchestration at different timescales. However, it focused on network service chains, without capturing all the intricate requirements and complex nature of MM sessions.

\emph{MuMeNet} fills this gap by $(i)$ embedding latency-annotated SGs onto a live packet topology, $(ii)$ tracking audio, video, and haptic objects separately to compute inter-stream skew at micro-second resolution, and $(iii)$ injecting bursty audience events that reshape traffic in real time. These capabilities enable faithful evaluation of orchestration algorithms under the exact timing constraints of musical co-presence.

\section{Reference Scenario: Audience Interaction in pre-recorded Live Concerts}\label{sec:musmopt}

\subsection{Reference Scenario Description}\label{subsec:SGMM}
We consider a reference MM scenario that entails a VR concert platform for distributed audiences using pre-recorded music and musicians' avatars on stage.  
The pre-recorded concert 
encompasses multitrack studio stems (audio), synchronized high-resolution video, lighting cues, and stage metadata, forming 
a passive immersive scene. Users, represented as audience avatars, can then participate and interact in real time, 
without any geographical constraints. 

We model this scenario as a SG $\mathcal{M=(N,K)}$, where vertices identify the required components for real-time interactive communications in the virtual concert, and edges their interconnections. 
A component in the SG belongs to one of these sets:
\begin{itemize}
    \item \emph{Producers} ($\mathcal{N}^s$): The entities generating continuous flows of data;
    \item \emph{Processors} ($\mathcal{N}^p$): The service functions responsible for processing the data generated by the producers, possibly receiving information also from databases;
    \item \emph{Consumers} ($\mathcal{N}^{d}$): The entities receiving and consuming data computed by the processors.
\end{itemize}

\begin{figure}[t]
    \centering
    \includegraphics[width=0.99\linewidth]{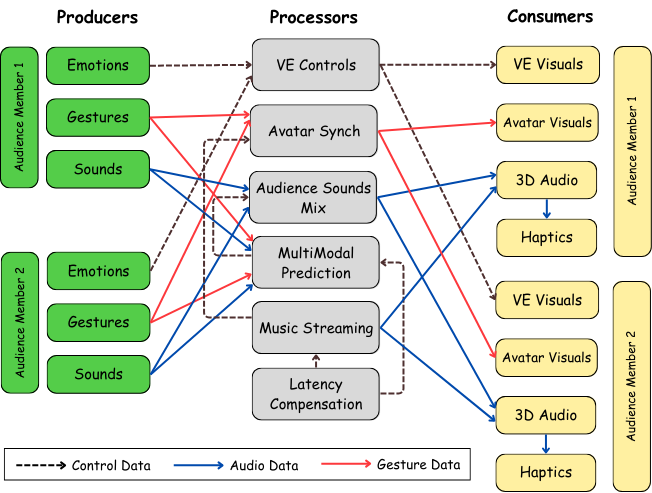}
    \caption{Graph representation of the modeled components and their connections of the reference SG given two users.}
    \label{fig:sg}
\end{figure}

Each user 
is represented in the graph by its \emph{producer} and \emph{consumer} components. Fig.~\ref{fig:sg} illustrates the SG for the described scenario, assuming two users.

The producers are the following:
\begin{itemize}
    \item \emph{Sounds:} This component tracks, via the microphone embedded in the Head-Mounted Display (HMD), the voice and contextual sounds generated by the audience member (e.g., while cheering, clapping), and transmits them onto the network;
    \item \emph{Gestures:} This component tracks the movements of the audience member (typically the position and rotation of head and hands), and transmits them onto the network at fixed intervals;
    \item \emph{Emotions:} This component retrieves in real-time, from biometric signals (e.g., EEG, heart rate), the emotional state of the audience member (as a class of 4 basic emotions as reported in \cite{emotions}), and transmits it onto the network at fixed intervals.
\end{itemize}

The processors are as follows:
\begin{itemize}
    \item \emph{Virtual Environment (VE) Controls:} This processor takes as input the emotional state of each audience member and produces as output (e.g., via majority voting schemes) a set of control messages for changing the parameters of the VE (e.g., brightness, contrast);
    \item \emph{Avatars Synchronizer:} This processor takes as input the gestures of audience members and uses them to control the embodiment of their corresponding avatar, synchronized across all instances of the multi-player VR application. 
    This processor also receives as input the data from the Multimodal Prediction module;
    \item \emph{Audience Sounds Mix:} This processor ingests the audio streams from every audience member. It blends the sounds of avatars positioned at a considerable distance from the listener’s own avatar into one mixed track, while sending to the listener the individual sound streams of up to ten nearby avatars (if their distance from the user is compatible with perceptual constraints). This processor also receives as input the data from the Multimodal Prediction module;
    \item \emph{Multimodal Prediction:} This module takes as input the data related to gestures, sounds and delay compensation, and produces predictions for gestures and sounds (via ML methods based on previous data) in order to cope with the heterogeneous latencies that would prevent audio-visual synchronization;
    \item \emph{Music Streaming:} This processor takes as input a database of compensation delays and uses them to stream music content to the connected audience members, each with a different delay. This ensures all geographically displaced audience members receive the music stream at the same time, regardless of the latency introduced by the network;
    \item \emph{Latency Compensation:} This processor implements a network controller that updates a database of network-related information used to compute compensation delays for the Music Streaming and Multimodal Prediction modules.
\end{itemize}

The consumers are as follows:
\begin{itemize}
    \item \emph{VE Visuals:} This component runs the multi-player VR application that updates the parameters of the VE based on the control messages received from the VE Controls;
    \item \emph{Avatars Visuals:} This component runs the multi-player VR application that updates the movements of the audience avatars based on the control messages received from the Avatars Synch;
    \item \emph{3D Audio:} This component receives as input the position of the closest avatars surrounding the avatar of the audience member and delivers to them their sounds spatialized according to 3D audio methods, along with the mix of the sounds of the farther avatars;
    \item \emph{Haptics:} This component provides a tactile representation of the music and other sounds, based on the audio signal produced by the 3D Audio device.
\end{itemize}

\subsection{Graph-Based Service and Network Modeling}

We leverage the CNFlow modeling and optimization framework \cite{CNFLOW} to propose a graph-based model for MM services and networks. Unlike VNE, which views a service as a collection of point-to-point demands without information awareness, i.e., without actual knowledge of the information they carry, CNFlow allows modeling information-aware service placement, routing, and resource allocation problem as a single, unified information-flow problem on an augmented cloud-network graph (CNG).

The CNG modeling the physical infrastructure not only includes communication links and nodes, but also available compute and storage resources at those nodes. Network nodes (e.g., user device, edge server, cloud data center) are characterized by their computational capacity to host and run service functions and possibly storing content, and network links are characterized by their communication capacity and propagation delay.

On the service side, the SG is an information-aware graph that captures the structure of the application's data flows and processing requirements. A SG is typically a DAG where \emph{vertices} represent service functions, e.g., audio analysis, mixing, spatial rendering, sensor processing, and \emph{edges} represent data streams, e.g., audio signals, control messages, flowing between functions or destined to end-users. Importantly, SGs in this graph modeling, include information attributes for each stream: for example, an audio stream might be one information object with a certain content,  bitrate, and latency requirement. In particular, we differentiate between commodities (edges in the SG) and information objects (attributes that determine the information carried by each commodity). 
A given \emph{information object} (say, a particular musician's audio stream) may be carried by multiple commodities, each indicating the need to receive that same information by multiple destinations (all other performers and audience members who should receive that audio).

\begin{figure}
    \centering
    \includegraphics[width=0.99\linewidth]{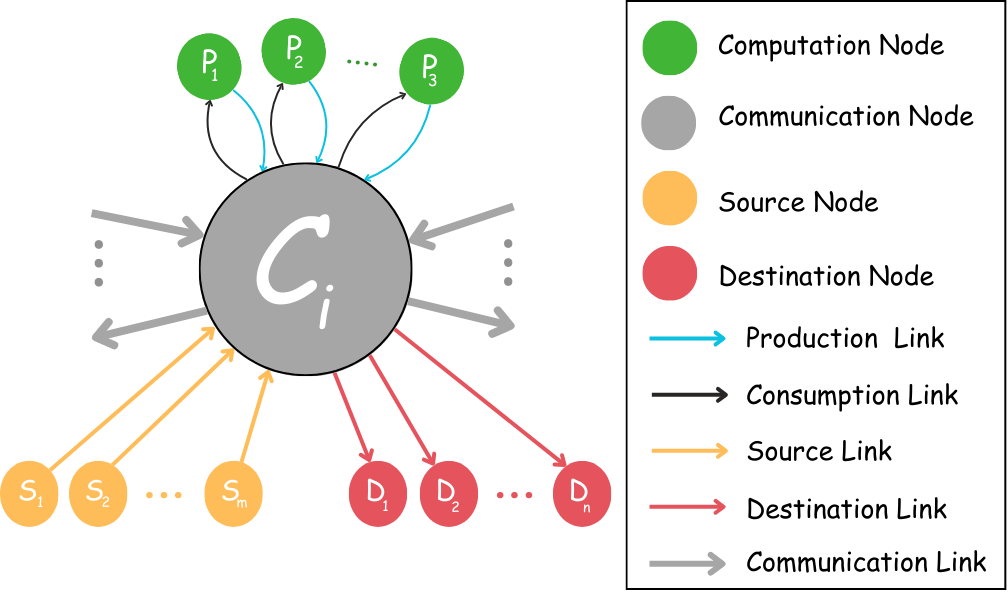}
    \caption{Augmented cloud-network graph. Gray edges represent network links indicating the availability of communication resources for transmitting information between nodes, yellow for producing data (source), red for consuming data (destination), blue and black are computation links representing producing and consuming capabilities, respectively.}
    \label{fig:node}
\end{figure}

\paragraph*{Cloud-Network Graph Model}\label{subsec:CNG} 
We model the physical infrastructure as a CNG $\mathcal{G}= (\mathcal{V},\mathcal{E})$, where vertices represent network nodes (e.g., core nodes, edge cloud nodes) and edges represent links between computing locations. Each node $i\in\mathcal V$ is augmented as illustrated in Fig.~\ref{fig:node}, to take into account their heterogeneous capabilities.
In the figure,  $\mathcal{V}^c$, $\mathcal{V}^s$, $\mathcal{V}^{p}$, and $\mathcal{V}^{d}$, are used to model the \emph{communication} (where messages are transmitted over the network), \emph{source} (where data is generated), \emph{computation} (where data is processed), and \emph{destination} (where data reaches its final destination) nodes, respectively. The resulting sets of nodes and links are denoted as: $\mathcal{V}=\mathcal{V}^{c} \cup \mathcal{V}^{s} \cup \mathcal{V}^{p} \cup \mathcal{V}^{d}$, and $\mathcal{E}=\mathcal{E}^{c} \cup \mathcal{E}^{s} \cup \mathcal{E}^{p} \cup \mathcal{E}^{d}$, denoting the set of communication links, source links, computation links, and destination links, respectively. In addition, computation links can either be a \emph{computation in} link $\mathcal{E}^{p_{in}} \subset \mathcal{E}^{p}$ (shown in black) or a \emph{computation out} link $\mathcal{E}^{p_{out}} \subset \mathcal{E}^{p}$ (shown in blue) \cite{idago}.

Each link $(i,j) \in \mathcal{E}$ is characterized by its capacity $c_{ij}$ and cost $w_{ij}$ parameters. In particular, for each $(i,j) \in \mathcal{E}^{c}$, $c_{ij}$ and $w_{ij}$ denote the capacity in communication flow units (e.g., bps) and the cost per unit flow, respectively.
Analogously, for each computation link $(i,j) \in \mathcal{E}^{p}$, $c_{ij}$ denotes the capacity in computation flow units. However, the unit of this value depends on the type of link: for $(i,j) \in \mathcal{E}^{p_{in}}$, $c_{ij}$ represents memory resources (e.g., RAM), hence, the corresponding value is expressed in bits. On the other hand, if $(i,j) \in \mathcal{E}^{p_{out}}$, $c_{ij}$ represents processing resources (e.g., CPU), it is  measured in FLOPs. Source and destination links $\mathcal{E}^{s}$, $\mathcal{E}^{d}$ are assumed to have zero cost and high-enough capacity, acting as network ingress and egress points, respectively. 

\begin{figure}
    \centering
    \includegraphics[width=0.99\linewidth]{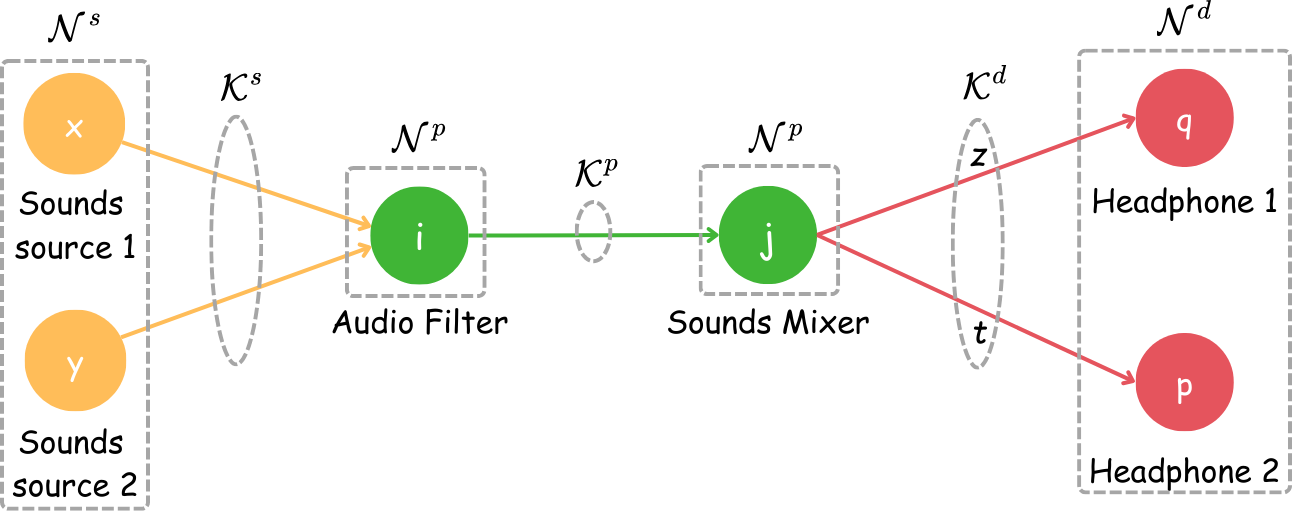}
    \caption{Example of a SG, where edges represent data streams (commodities) and vertices service functions.}
    \label{fig:transition}
\end{figure}

\paragraph*{Service Graph model}\label{subsec:SG} A SG for a MM application is a graph $\mathcal{M}=(\mathcal{N}, \mathcal{K})$, where vertices represent service functions (e.g., a user sounds decoder) and edges corresponding data streams (or commodities), as shown in Fig.~\ref{fig:transition}.

A function $n \in \mathcal{N}$ can either be a \emph{source node} (a producer of data), a \emph{processing node} (a processor data), or a \emph{destination node} (a consumer of data). Similarly, a commodity in $\mathcal K$ can either be a \emph{source commodity}, \emph{processing commodity}, or \emph{destination commodity}. Analogously, $\mathcal{N} = \mathcal{N}^{s} \cup \mathcal{N}^{p} \cup \mathcal{N}^{d}$, and consequently $\mathcal{K} = \mathcal{K}^{s} \cup \mathcal{K}^{p} \cup \mathcal{K}^{d}$. An commodity $k \equiv (i,j) \in \mathcal{K}^{p} \cup\mathcal{K}^{p} $ represents a commodity produced by function $i \in \mathcal{N}^{s}\cup\mathcal{N}^{p}$, and consumed by function $j \in \mathcal{N}^{p}\cup \mathcal{N}^{d}$. We denote by $\mathcal{X}(k)$ the set of required input commodities to produce commodity $k$. For example, for the commodity $(i,j)$ shown in Fig.~\ref{fig:transition}, $\mathcal{X}((i,j))$ is the set of the two incoming source commodities $\mathcal{K}^s$. On the other hand, $\mathcal{X}(z)$ consists only of commodity $(i,j)$. Moreover, we also denote by $s(k) \in \mathcal{V}^{s}$ the node hosting the function producing source commodity $k \in \mathcal{K}^{s}$, $d(k) \in \mathcal{V}^{d}$  the node hosting the function consuming destination commodity $k\in \mathcal{K}^{d}$. 

In $\mathcal{M}$, each commodity is characterized by a 
rate requirement $R^{k}_{ij}$ 
denoting the required rate of commodity $k\in \mathcal{K}$ when it goes over link $(i,j) \in \mathcal{E}$. Note that the rate of $k$ will depend on the type of link $(i,j)$ it traverses. We use $R^{k}_{prod}$, $R^{k}_{comm}$, and $R^{k}_{cons}$ to denote the rate of commodity $k$ when it goes over a production link, communication link, or consumption link, respectively. 

Finally, as described in \cite{idago}, one of the most important aspects for efficient representation of real-time data-intensive applications is the concept of information awareness. We define the set of information objects $\mathcal{O}$, and the surjective \emph{information mapping function} $g:\mathcal{K}\to\mathcal{O}$ to indicate the object identifier $o\in \mathcal{O}$ associated with each commodity $k$. This mapping function is key to allow the overlapping of commodity flows carrying the same information.

\section{\textsc{MuMeNet}: Musical Metaverse Network Simulator}\label{sec:simulator}
We now present MuMeNet, our proposed discrete-event simulator for the MM. MuMeNet allows researchers to $(i)$ develop flexible algorithm to embed MM dynamic SGs on arbitrary network topologies with real-world parameters, $(ii)$ stream fine-grained traffic commodities through those embeddings under realistic link-level impairments, and $(iii)$ measure, study, and analyze end-to-end musical-quality, QoE, and QoS for users with a large set of extensible evaluation metrics.

\begin{figure}
    \centering
    \includegraphics[width=0.95\linewidth]{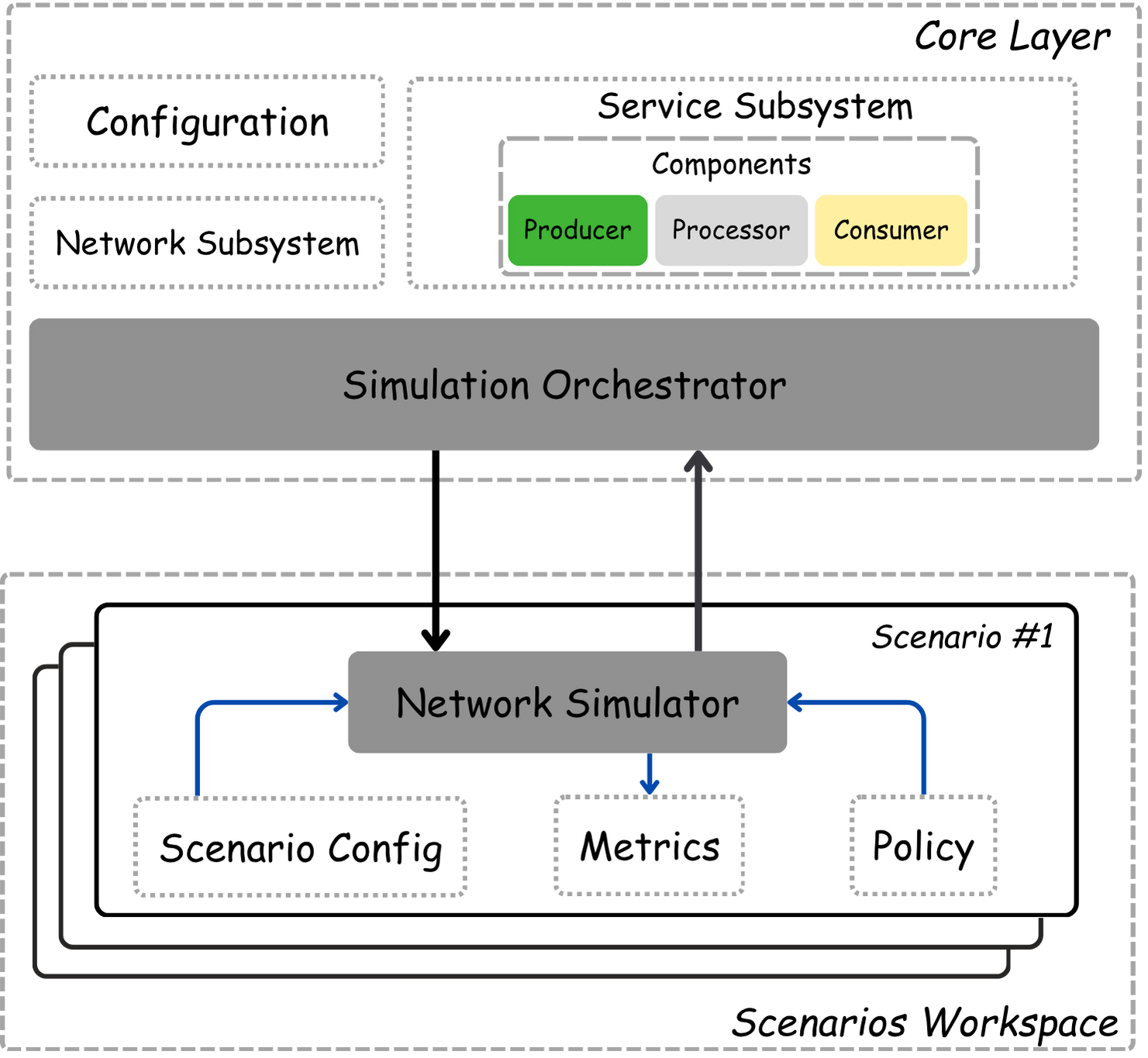}
    \caption{MuMetNet Design Architecture: The top part shows the \emph{Core Layer} responsible for general orchestration and management, while the bottom part shows \emph{Scenarios Workspace} where event-driven experiments are performed.}
    \label{fig:arch}
\end{figure}

Fig.~\ref{fig:arch} depicts the architectural design of MuMeNet, which consists of two main parts: the \emph{core layer} and the \emph{scenarios workspace}. 

\subsection{Core Layer}
The Core Layer is the invariant runtime backbone of MuMeNet. It collects every (simulation) service that must remain stable across experiments (e.g., configuration management, network modeling, component life-cycle control, and the discrete-event orchestrator) into a single, version-controlled module. It bundles the following four subsystems:

\paragraph*{Simulator Configuration} This subsystem provides the authoritative source for all runtime parameters needed by the simulator. At start-up, the engine instantiates a \emph{BaseConfig} object that pulls a version-controlled YAML/JSON file with default values (e.g., simulation horizon, time step, random seeds, logging levels, and directories), validates every field against a schema and reports any mismatch, then archives the resolved configuration alongside the run outputs to preserve full reproducibility.

This architectural approach facilitates extensibility through inheritance-based configuration customization. Researchers can derive domain-specific configuration classes from the \emph{BaseConfig} base class within their respective scenario directories. Such derived classes retain the capability to override existing attributes or introduce additional parameters while preserving the underlying validation and persistence mechanisms. Upon initialization, the simulation engine maintains an immutable reference to the configuration object, which serves as the canonical parameter source for all core subsystems, including the network model, metrics collection framework, and policy plug-in architecture. This design pattern ensures centralized parameter management with strong type safety guarantees while simultaneously enabling scenario-specific customization without requiring modifications to the core system implementation.

\paragraph*{Network Subsystem}
This subsystem is the bridge between the abstract network graph template and the event traffic that flows through the simulator at run time. It performs two high-level roles:
\begin{itemize}
    \item Graph Construction: Assembling a topology whose nodes and links carry detailed resource attributes (e.g., packet loss, jitter, propagation delay). Moreover, a user can easily add some other dimensions.
    \item Topological Consistency Checking: Ensuring that the constructed topology obeys domain-specific constraints so that no unrealistic artifacts bias the results. 
\end{itemize}

Note that at the start of every run, the network subsystem ingests either a ready-made template derived from real deployments or a synthetic topology. 

\paragraph*{Service Subsystem} A SG is realized as an ordered \emph{set of components} (see Sec.~\ref{sec:musmopt}). Each component encapsulates the internal logic of a single service-graph node and is instantiated as one of three canonical roles (Fig.~\ref{fig:sg} shows how these roles combine to form the pre-recorded-concert template). 
Custom behavior is added by subclassing the role-specific bases (Producer, Processor, or Consumer), while the shared interface keeps every extension drop-in compatible with the framework. A component in the simulator passes through a set of different phases. First, the developer instantiates the object, during this phase the simulator validates the provided metadata (e.g., I/O rates, state size, latency budget) against domain constraints. Second, the component is added to the global registry, making it discoverable by other elements. Finally, after having all required components successfully registered, the simulator connects them via provisional data-flows and exchanges lightweight \lq \lq dummy'' messages to verify type compatibility, buffer sizes, and back-pressure signals. Any mismatch triggers a descriptive error before the actual experiment begins.


\paragraph*{Simulation Orchestrator}
The Orchestrator constitutes the control plane of the simulator.
Its design is deliberately centered on a discrete-event scheduling paradigm, because audience interaction workloads exhibit highly uneven traffic bursts and tight latency budgets that are best captured by event-driven, rather than fixed-timestep execution.

The orchestrator fulfills the following responsibilities:
\begin{itemize}
    \item Global Time Monitoring: It moves the simulation clock to the timestamp of the next event in the queue.
    \item Event Dispatcher: It invokes the \emph{callback} attached to each event, thereby triggering computation, communication, or rendering inside the appropriate component. The event also accepts a set of optional arguments if required.
    \item State Monitor and Update: It updates link capacities, queue lengths, and component states after every callback, guaranteeing that subsequent events observe a consistent world view.
    \item Variability Engine: It injects stochastic effects (propagation-delay distributions, packet-loss bursts, device jitter) at the precise moment they materialize, this information is initially provided by the user in the configuration files.
\end{itemize}

Every occurrence in the simulator such as, e.g., sending a packet, completing a CPU task, finishing an audio render, is encoded as an \textit{event object} with three immutable fields: \emph{unique ID}, \emph{timestamp} (at which the event must start executing), and \emph{callback} (the function to execute).

\begin{figure}
    \centering
    \includegraphics[width=0.9\linewidth]{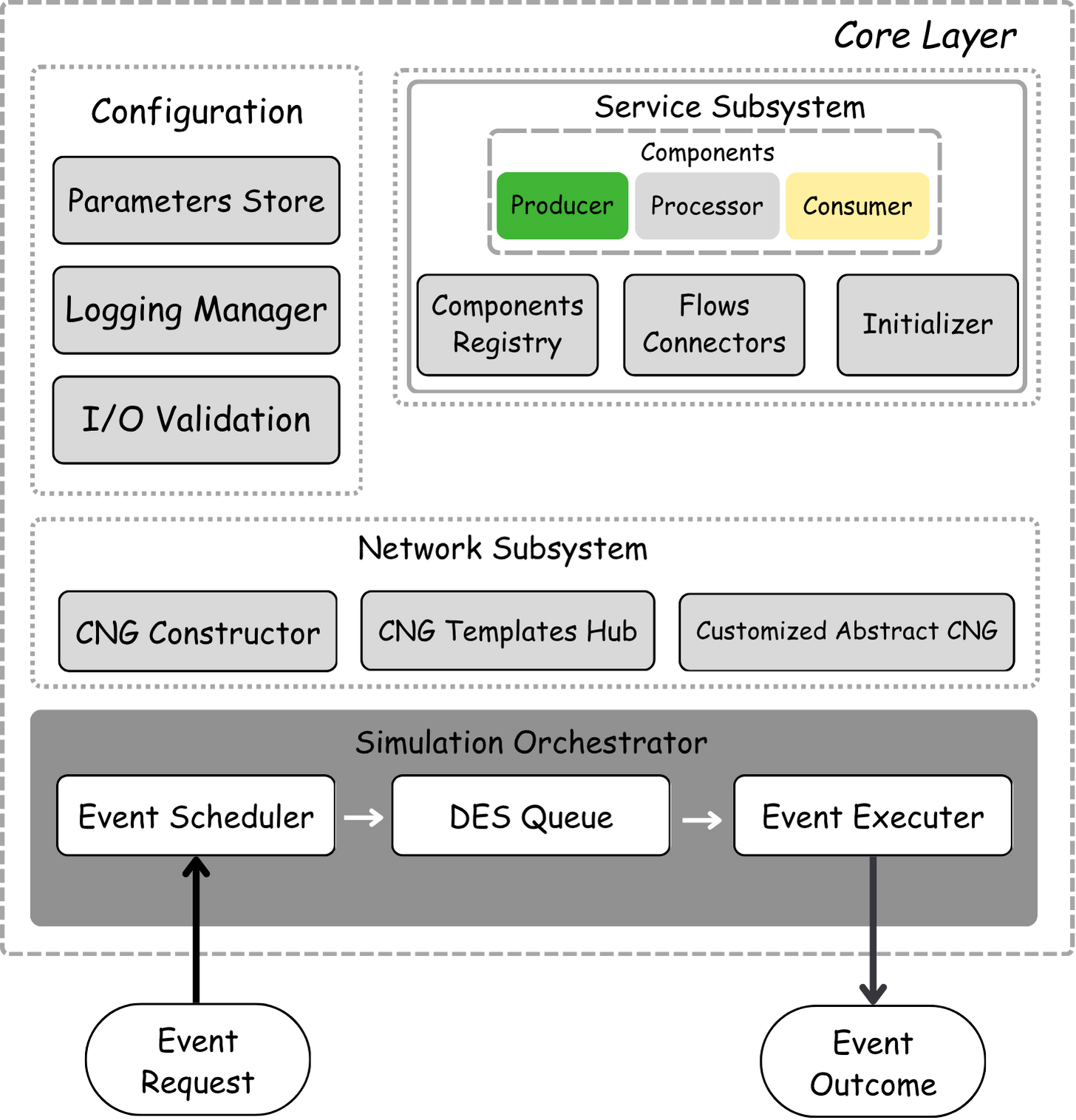}
    \caption{Core Engine Architecture of MuMeNet.}
    \label{fig:detailed}
\end{figure}

Fig.~\ref{fig:detailed} provides a more in-depth view of the core engine, where each module/subsystem is shown along with its parts.

\subsection{Scenarios Workspace}
This workspace allows users to create their customized scenarios. 
Fig.~\ref{fig:arch} shows the general architectural design of a scenario inside the workspace. 
Four modules compose the workspace, as detailed below.
\paragraph*{Scenario Configuration Module}
This module can extend the set of initial configuration parameters defined in the \emph{core} module. The reason behind this design is to ensure separability: only the necessary parameters are called, otherwise they remain set to default. In addition to this, we store the necessary scenario configuration in this module. We share with the user the set of possible parameters to consider for the pre-defined scenarios; however, if the user wants to create a completely customized scenario out of the provided template, this may require additional steps as values have to be defined (and validated) first in the core part.

\paragraph*{Policy Module}
This module contains a set of abstract classes and methods to be extended for easy design of a customized policy. The policy’s objective is to determine placement, routing, and resource allocation for the incoming SGs over the CNG, following static and dynamic approaches. 

\paragraph*{Metrics Module}
This module collects real-time metrics of all components, monitors the changes of their statuses, and updates the list of metrics. This list depends on the use case; however, users can easily extend the base classes provided in this module to create any customized metric along with the interval of measuring or the set of components to use for collecting results.

\paragraph*{Network Simulation Module}
The scenario's key subsystem bridges the scenario with the core layer: it runs the orchestrator's network-level callbacks and returns updates or event requests, acting as the simulator’s central communication hub, where all connections inside the simulator meet, from the CNG to the SG, and where metrics are collected before being sent to the \lq \lq Metrics'' module.

\section{Problem Formulation and MILP Model}
\label{sec:ilp}

The processors embedding problem encompasses the placement of processors and routing traffic flows within the SG over the CNG, given user locations. It is formulated as follows:
\emph{given} a CNG $\mathcal{G}=(\mathcal{V},\mathcal{E})$ with a set of compute-capable nodes interconnected by links with bandwidth and propagation delay; a set of users connected to specific access nodes; and a SG $\mathcal{M}=(\mathcal{N},\mathcal{K})$ representing the MM service pipeline, \emph{decide} the placement of each processor in $\mathcal N^p$ and the route for each commodity in $\mathcal K$,
with the \emph{objective} of either minimizing the sum of costs over utilized links, or minimizing the maximum link utilization,
\emph{constrained} by flow conservation, flow-chaining, source initialization, destination collection, commodity to information mapping, capacity bounds, and end-to-end latency bounds. Sets, list of variables, and list of parameters are reported in Table~\ref{tab:sets}.

\paragraph*{Variables} We define two groups of variables:
\begin{enumerate}
    \item \emph{Virtual Commodity Flows $\{ f_{ij}^k\}$}: Adimensional binary variables indicating whether commodity $k\in \mathcal{K}$ goes (i.e., is transmitted, processed, or stored) over link $(i,j) \in \mathcal{E}$.
    \item \emph{Actual Information Flows $\{\mu^o_{ij}\}$} and \emph{$\{\mu_{ij}\}$ }: Real variables indicating the \emph{amount} of information flow associated with object $o \in \mathcal{O}$ and the \emph{total} information flow, respectively, going over link $(i,j) \in \mathcal{E}$. 
\end{enumerate}

\paragraph*{Objectives} 
We consider two different objective functions in our analysis:
\begin{enumerate}
    \item Minimum-Cost (\emph{MinCost}): This objective seeks the minimum cost feasible embedding by minimizing the sum of costs (total cloud-network resource cost) over utilized link resources, as shown in Eq.\ref{eq:mincost}.
    \begin{equation}
        \sum_{(i,j) \in\mathcal{E}}\mu_{ij}w_{ij}
        \label{eq:mincost}
    \end{equation}

    \item Load Balancing (\emph{LB}): This objective minimizes the maximum utilization over links (i.e., maximum saturated link), with the aim of ensuring a balanced resources allocation in the network. This can be modeled as:
    \begin{equation}
        \text{min}  
        \Bigr[ \max_{(i,j)\in\mathcal{E}} U_{ij} \Bigr] ; U_{ij} = \frac{\mu_{ij}}{c_{ij}}
    \label{eq:lb}
    \end{equation}

    where $U_{ij}$ is the utilization (or load) of link $(i,j) \in \mathcal{E}$. Note that, since the nested min-max breaks linearity, we linearize it using the standard epigraph trick. Specifically, we use a single auxiliary variable $\mathbf{Z}$, resulting in an additional constraint. The final form becomes:
\begin{align}
    \text{min} & \quad \mathbf{Z} \\
    \text{s.t.} & \quad \mu_{ij} \leq \mathbf{Z} \cdot c_{ij} \\
    \quad & \quad \mathbf{Z} \in \mathbb{R}^{+} \\
    \quad & \quad \text{Constraints} \quad \eqref{eq:b} \text{-} \eqref{eq:l} 
\end{align}
\end{enumerate}

\paragraph*{Constraints}
{\small
\begin{align}
    \text{s.t.} \quad 
    & \sum_{j \in \delta^{-}(i)} f_{ji}^{k} = \sum_{j \in \delta^{+}(i)} f_{ij}^{k}
    \quad \forall i \in \mathcal{V}, k \in \mathcal{K} \label{eq:b} \\
    & f_{ij}^{k} = 
    \left\{
    \begin{array}{ll}
        f_{ji}^{l} & \forall k \in \mathcal{K}^{p} \\
        0 & \text{otherwise}
    \end{array}
    \right.
    \forall l \in \mathcal{X}(k), i \in \mathcal{V}^{p} ,j \in \mathcal{\delta}^{+}(i)
    \label{eq:c} \\
    & f_{ij}^{k} = 
    \left\{
    \begin{array}{ll}
        1 & \forall k \in \mathcal{K}^{s} \\
        0 & \text{otherwise}
    \end{array}
    \right.
    \quad \forall i = s(k), j \in \mathcal{\delta}^{+}(i)
    \label{eq:d} \\
    & f_{ij}^{k} = 
    \left\{
    \begin{array}{ll}
        1 & \forall k \in \mathcal{K}^{d} \\
        0 & \text{otherwise}
    \end{array}
    \right.
    \quad \forall j = d(k), i \in \mathcal{\delta}^{-}(j)
    \label{eq:e} \\
    & f^{k}_{ij} R^{k}_{ij} \le \mu^{o}_{ij}
    \quad \forall (i,j) \in \mathcal{E}, k \in \mathcal{K}, o = g(k)
    \label{eq:f} \\
    & \sum_{o\in \mathcal{O}} \mu^{o}_{ij} \le \mu_{ij} \le c_{ij}
    \quad \forall (i,j) \in \mathcal{E}
    \label{eq:g} \\
    & l^{k} = \sum_{(i,j) \in \mathcal{E}} l^{k}_{ij} f^{k}_{ij}
    \quad \forall k \in \mathcal{K}
    \label{eq:h} \\
    & l_{T}^{k} = l^{k}
    \quad \forall k \in \mathcal{K}^s
    \label{eq:i} \\
    & l_{T}^{k} \ge l^{k} + l^{l}_{T}
    \quad \forall k \in \mathcal{K}^p \cap \mathcal{K}^d , l \in \mathcal{X}(k) 
    \label{eq:j} \\
    & l_{T}^{k} \le L^{k}
    \quad \forall k \in \mathcal{K}^d
    \label{eq:k} \\
    & \begin{array}{l}
        f_{ij}^{k} \in \{0,1\}, \mu^{o}_{ij} \in \mathbb{R}^{+}, \mu_{ij} \in \mathbb{R}^{+}, l^{k} \in \mathbb{R}^{+}, l^{k}_{T} \in \mathbb{R}^{+}
    \\ \forall (i,j) \in \mathcal{E}, k\in \mathcal{K}, o\in \mathcal{O}
    \label{eq:l}
    \end{array}
\end{align}
}

\begin{table}[tbp]
\caption{System model sets, parameters and variables}
\label{tab:sets}
\centering
\renewcommand{\arraystretch}{0.9}%
\begin{tabular}{|lp{6cm}|}
\hline
\emph{Sets} & \emph{Description} \\ 
\hline
$\mathcal{G=(V,E})$ & Cloud-Network graph  \\
$\mathcal{V}^c;\mathcal{V}^s;\mathcal{V}^p;\mathcal{V}^d$ & Communication nodes; source nodes; computation nodes; destination nodes \\
$\mathcal{E}^c;\mathcal{E}^s;\mathcal{E}^p;\mathcal{E}^d$ & Communication links; source links; computation links; destination links \\ 
$\mathcal{E}^{p_{in}};\mathcal{E}^{p_{out}}$ & Computation in links (storage resources); Computation out links (processing resources) \\ 
$\mathcal{\delta}^+(i); \mathcal{\delta}^-(i)$ & Incoming and outgoing neighbors of node $i \in \mathcal{V}$ \\ 
$\mathcal{M=(N,K)}$ & Service graph DAG \\ 
$\mathcal{N}^s;\mathcal{N}^d;\mathcal{N}^p$ & Source functions; destination functions; processing functions \\ 
$\mathcal{K}^s;\mathcal{K}^d;\mathcal{K}^p$ & Source commodities; destination commodities; processing commodities \\ 
$\mathcal{X}(k)$ & Set of input commodities required to produce commodity $k$ \\ 
$s(k);d(k)$ & Source node hosting the function producing commodity $k \in \mathcal{K}^s$; Destination node hosting the function consuming commodity $k \in \mathcal{K}^d$. \\ 
\hline
\emph{Parameters} & \emph{Description} \\ 
\hline

$c_{ij} \in \mathbb{R}^{+}$ & Capacity of link $(i,j)$ \\
$w_{ij} \in \mathbb{R}^{+}$ & Cost of link $(i,j)$ \\
$U_{ij} \in [0,1]$ & Load of link $(i,j)$ \\
$R^k_{ij} \in \mathbb{R}^{+}$ & Rate of $k \in \mathcal{K}$ when it goes over link $(i,j)$ \\ 
$l_{ij}^{k} \in \mathbb{R}^{+}$ & Latency to transmit or process a unit of $k$ over link $(i,j)$ \\
$l^{k}\in \mathbb{R}^{+}$ & Local latency of commodity $k$ \\
$L_{T}^{k} \in \mathbb{R}^{+}$ & Cumulative latency of commodity $k$ \\
$L^{k} \in \mathbb{R}^{+}$ & Maximum service latency associated with destination commodity $k$\\
$\mathbf{Z} \in [0,1]$ & Utilization upper bound \\
\hline
\emph{Variables} & \emph{Description} \\
\hline
$f_{ij}^k \in \{0,1\}$ & Virtual commodity flow \\
$\mu_{ij}^o \in \mathbb{R}^{+}$ & Actual information flow \\
$\mu_{ij} \in \mathbb{R}^{+}$ & Total  information flow  \\

\hline
\end{tabular}
\end{table}

Constr.~\ref{eq:b} generalizes communication, computation, and storage flow conservation, requiring the total incoming flow to a given communication node $i \in \mathcal{V}$ for a given commodity $k \in \mathcal{K}$ to be equal to the total outgoing flow from node $i$ for commodity $k$. 
Constr.~\ref{eq:c} enforces the flow chaining condition, i.e., a computation node $i \in \mathcal{V}^{p}$ can produce commodity $k \in \mathcal{K}$ only if all required input commodities $l \in \mathcal{X}(k)$ are available at its input. For instance, generating an enhanced video ($k$) requires the decoded video ($l$) as input.
Constr.~\ref{eq:d} and \ref{eq:e} define the source and destination conditions, ensuring that each commodity $k \in \mathcal{K}^s$ is generated at its unique source node $s(k)$ and delivered to its destination $d(k)$. 
Constr.~\ref{eq:f} ensures that overlapping flows carrying the same information can share link capacity, reflecting the multicast nature of MM services. 
Recall that MM services involve real-time data streams required by multiple processing and/or destination functions. To capture this, the formulation allows virtual commodity flows corresponding to the same information object to overlap on a link $(i,j) \in \mathcal{E}$. This is achieved by weighting each flow variable with its rate requirement and summing only distinct information objects, not individual commodities.
Constr.~\ref{eq:g} limits the total information flow on each link $(i,j) \in \mathcal{E}$ to its available capacity. This total flow is computed by summing the sizes of all distinct information flows traversing the link. Constr.~\ref{eq:h}-\ref{eq:k} ensure end-to-end service latency constraints. Specifically, 
Constr.~\ref{eq:h} enforces that the local latency of commodity $k$, $l^k$, (i.e., the time taken to produce, deliver and consume a unit of commodity $k$) is the sum, on the links carrying commodity $k$, of the latency to transmit or process a unit of commodity $k$ over the given link, indicated by $l^{k}_{ij}$. Constr.~\ref{eq:i}-\ref{eq:j} computes the cumulative latency of commodity $k$, $l^{k}_{T}$, which represents the service latency that has been accumulated until the consumption of commodity $k$. Constr.~\ref{eq:i} ensures that the cumulative latency is equal or less than the local latency for all source commodities. Constr.~\ref{eq:j} computes the cumulative latency for all remaining commodities recursively by setting the cumulative latency of commodity $k$ to be larger than or equal to the local latency of commodity $k$ plus the cumulative latency of input commodity $l$, for all input commodities in $\mathcal{X}(k)$. Lastly, Constr.~\ref{eq:k} imposes the cumulative latency at each destination commodity to be no greater than the maximum allowed service latency $L^{k}$, while Constr.~\ref{eq:l} imposes the binary nature of commodity flow variables and the real positive nature of information flow and latency variables.


\section{Experimental Settings and Results}\label{sec:results}

\subsection{Experimental Settings}
We perform our evaluations considering two distinct synthetic cloud network graphs (CNG) generated as a binomial topology $G(n,\rho)$ with $n \in \{8,16\}$ with an edge-existence probability $\rho = 0.7$ \cite{erd1959random}. Link capacities and costs are drawn from a uniform distribution $\mathcal U(a,b)$, whose range of values is reported in Table~\ref{tab:params}. We consider the \emph{computation-node percentage}, i.e., the per-communication-node probability of hosting at least one computation node, to be equal to $0.7$ to reflect heterogeneous edge–core deployments. We also consider $N$ = 19 SGs (each SG has an audience number equal to $|\mathcal{U}| = N+1$ users). Table \ref{tab:SGEdgeRanges} reports the parameters used to generate the 19 SGs. Each SG follows the modeling shown in Fig.~\ref{fig:sg}.
We evaluate the MILP using the two objective functions described in Sec~\ref{sec:ilp}. Although our experiments highlight the trade-off between monetary cost and network resilience under different load conditions, the primary objective is to demonstrate the types of analyses that can be performed using MuMeNet. The reported values are obtained by aggregating results from 100 repeated runs for each parameter combination, with confidence intervals calculated at a 95\% confidence level ($\alpha = 0.05$).

\begin{table}[t]
\caption{Configuration Setup for Cloud Network Topologies}
\label{tab:params}
\centering
\renewcommand{\arraystretch}{0.9}%
\begin{tabular}{|l|c|}
\hline
\emph{Parameter} & \emph{Range of values} \\ 
\hline
Communication Links capacity & $\mathcal{U}(1000,3000)$ \\
\hline
Communication Links cost & $\mathcal{U}(10,30)$ \\
\hline
Computation in capacity (RAM) & $\mathcal{U}(1000,3000)$ \\
\hline
Computation in cost & $\mathcal{U}(10,25)$ \\
\hline
Computation out capacity (CPU) & $\mathcal{U}(1000,2000)$ \\
\hline
Computation out cost & $\mathcal{U}(10,25)$ \\
\hline
Computation Percentage & 0.7 \\
\hline
Number of computation nodes & $\mathcal{U}(1,3)$ \\
\hline
Density & 0.7 \\
\hline
\end{tabular}
\end{table}

\begin{table}[htbp]
\centering
\renewcommand{\arraystretch}{0.9}%
\caption{Considered values of production, communication and consumption rates assigned to each synthetic SG edge.}
\label{tab:SGEdgeRanges}
\scalebox{0.9}{%
\begin{tabular}{|c|c|c|c|c|}
\hline
\textbf{Pattern (src$\;\rightarrow\;$dst)} & $\boldsymbol{R_{\text{prod}}}$ & $\boldsymbol{R_{\text{comm}}}$ & $\boldsymbol{R_{\text{cons}}}$ \\ \hline

\multicolumn{4}{|c|}{\textbf{Source}} \\ \hline
$\_\,\rightarrow\,$VEControls               & $\mathcal{U}(3,10)$   & $\mathcal{U}(10,50)$ & $\mathcal{U}(10,50)$ \\ \hline
$\_\,\rightarrow\,$AvatarSynch              & $\mathcal{U}(50,70)$  & $\mathcal{U}(10,50)$ & $\mathcal{U}(10,50)$ \\ \hline
$\_\,\rightarrow\,$AudienceMix              & $\mathcal{U}(80,120)$ & $\mathcal{U}(10,50)$ & $\mathcal{U}(10,50)$ \\ \hline
MMPred$_{\text{gest}}\,\rightarrow\,$AvatarSynch  & $\mathcal{U}(50,70)$  & $\mathcal{U}(10,50)$ & $\mathcal{U}(10,50)$ \\ \hline
MMPred$_{\text{sound}}\,\rightarrow\,$AudienceMix & $\mathcal{U}(80,120)$ & $\mathcal{U}(10,50)$ & $\mathcal{U}(10,50)$ \\ \hline

\multicolumn{4}{|c|}{\textbf{Destination}} \\ \hline
VEControls$\,\rightarrow\,\_$   & $\mathcal{U}(10,50)$ & $\mathcal{U}(10,50)$ & $\mathcal{U}(3,10)$    \\ \hline
AvatarSynch$\,\rightarrow\,\_$  & $\mathcal{U}(10,50)$ & $\mathcal{U}(10,50)$ & $\mathcal{U}(50,70)$   \\ \hline
AudienceMix$\,\rightarrow\,\_$  & $\mathcal{U}(10,50)$ & $\mathcal{U}(10,50)$ & $\mathcal{U}(120,200)$ \\ \hline
Streaming$\,\rightarrow\,\_$    & $\mathcal{U}(10,50)$ & $\mathcal{U}(10,50)$ & $\mathcal{U}(120,200)$ \\ \hline

\multicolumn{4}{|c|}{\textbf{Processing}} \\ \hline
MMPred$\,\rightarrow\,$AvatarSynch & $\mathcal{U}(50,70)$  & $\mathcal{U}(30,70)$ & $\mathcal{U}(50,70)$  \\ \hline
MMPred$\,\rightarrow\,$AudienceMix & $\mathcal{U}(80,120)$ & $\mathcal{U}(30,40)$ & $\mathcal{U}(80,120)$ \\ \hline
LatComp$\,\rightarrow\,$Streaming  & $\mathcal{U}(3,10)$   & $\mathcal{U}(3,10)$  & $\mathcal{U}(3,10)$   \\ \hline
LatComp$\,\rightarrow\,$MMPred     & $\mathcal{U}(3,10)$   & $\mathcal{U}(3,10)$  & $\mathcal{U}(3,10)$   \\ \hline
\end{tabular}}
\end{table}

\subsection{Numerical Results}

\begin{figure}[tb]
    \centering
    \includegraphics[width=0.9\linewidth]{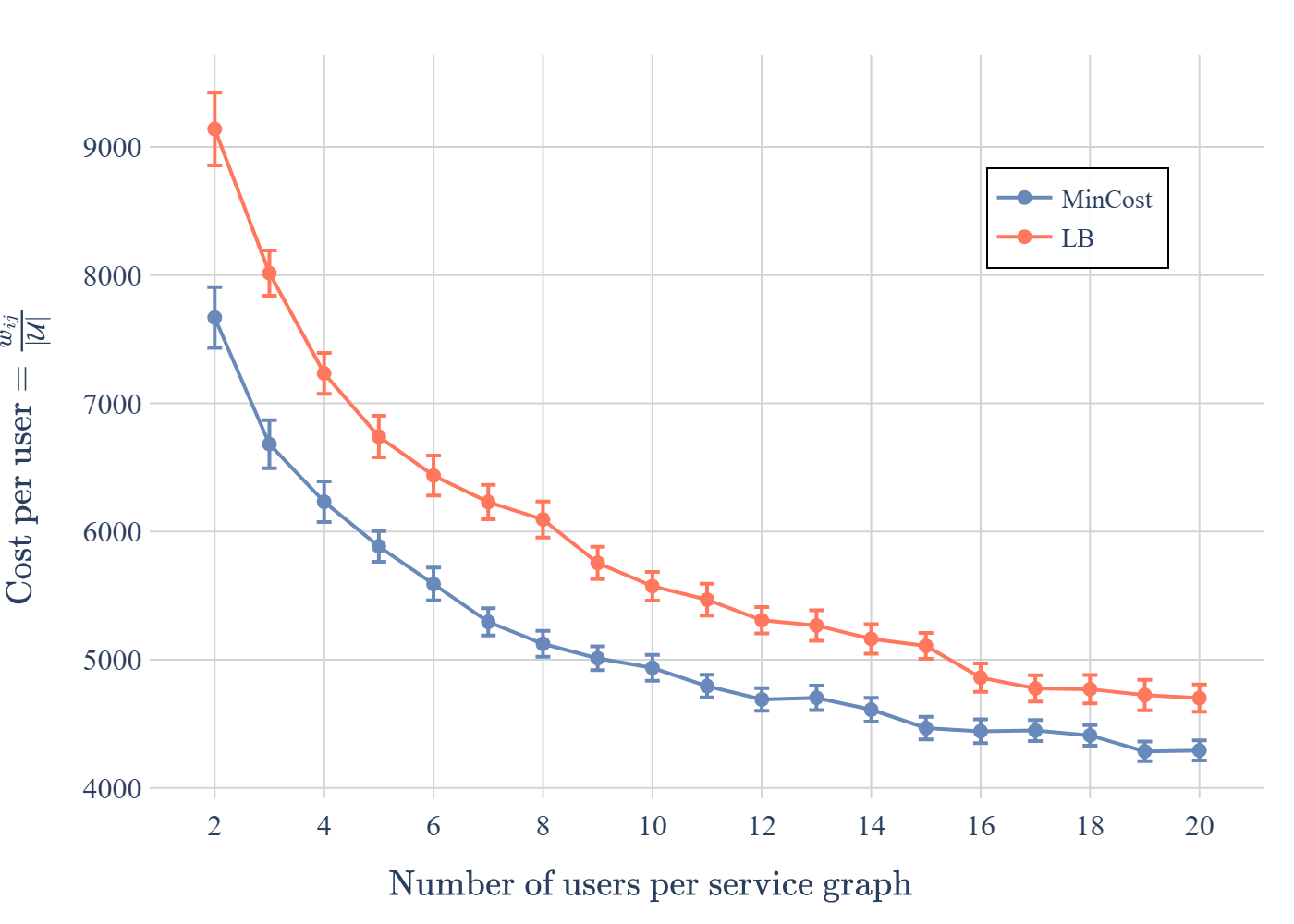}
    \caption{Cost per number of users for varying number of users in the SG for 8-node binomial network topology.}
    \label{fig:cng8cost}
    \vspace{-0.4cm}
\end{figure}

\begin{figure}[tb]
    \centering
    \includegraphics[width=0.9\linewidth]{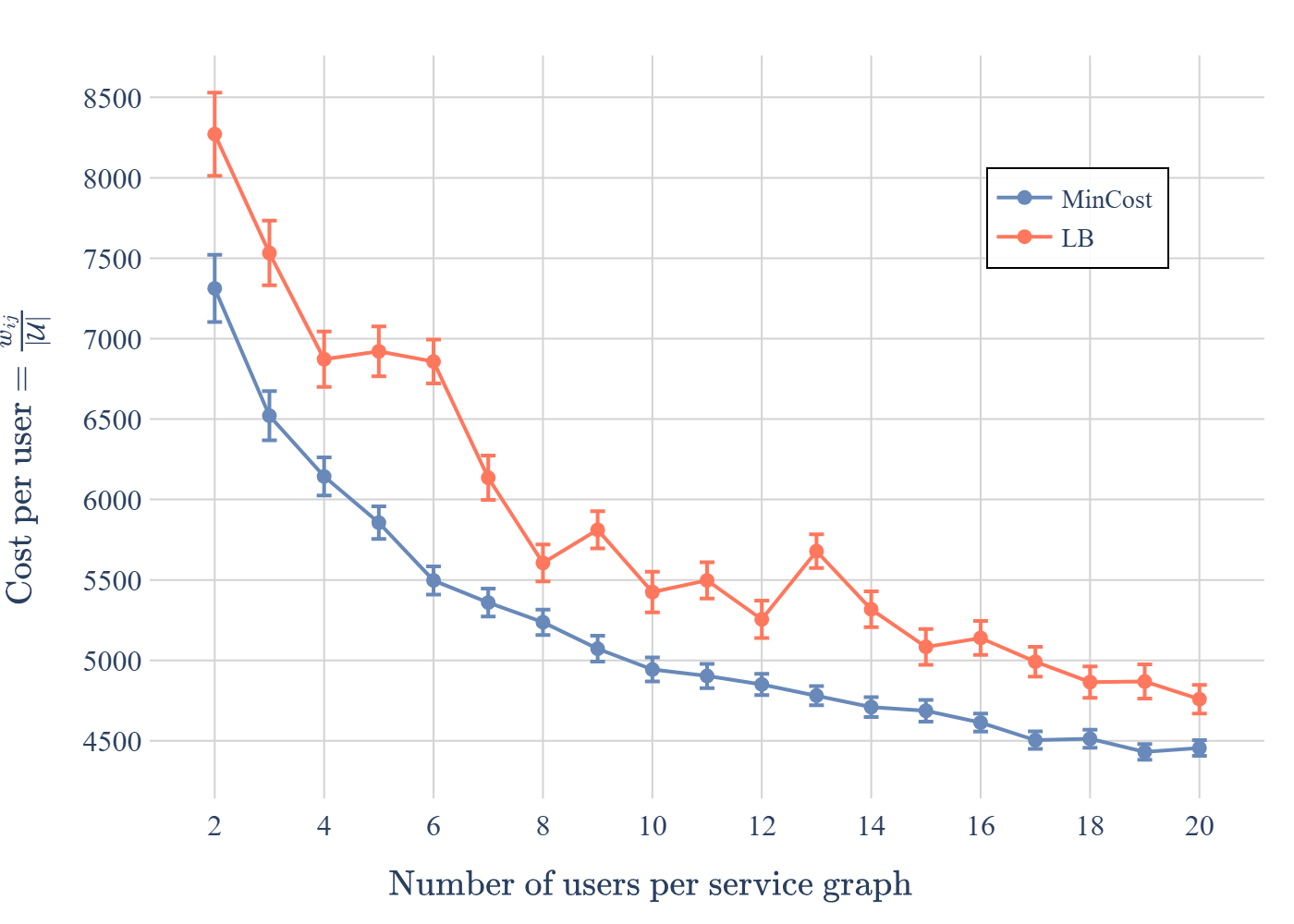}
    \caption{Cost per number of users for varying number of users in the SG for 16-node binomial network topology.}
    \label{fig:cng16cost}
\end{figure}

Figures ~\ref{fig:cng8cost} and \ref{fig:cng16cost} show the cost per user achieved by the MILP when employing the two different objective functions for varying numbers of users per SG for 8-node and 16-node CNG, respectively. In both cases, and for each objective, the cost per user shows a general downward trend, i.e., it decreases with more users per SG, with \emph{MinCost} achieving between 10\% to 15\% and between 10\% to 33\% less cost than \emph{LB} for the 8-node CNG and 16-node CNG, respectively\footnote{The \emph{LB} shows a fluctuating curve for the 16-node CNG case. This is due to the fact that the objective is insensitive to small increases in total price as long as peak utilization is reduced.}. This indicates that when more users share the same MM session, traffic can be aggregated and common data streams can be shared, driving the per-user cost down.

More specifically, with 2 users per SG, \emph{MinCost} achieves a cost of roughly 7700 units per user, whereas the \emph{LB} achieves a cost of user 9200 (18\% more). This surcharge arises as the \emph{LB} deliberately routes traffic to minimize peak utilization, even if it incurs additional monetary costs (i.e., even if it utilizes costlier links). 
Moreover, we further note that starting from 15 users per SG for both 8 and 16-node CNGs, the two curves begin to align more closely (the difference at 20 users is $\approx$ 350 units, or $\approx$ 10\%). This indicates that with more concurrent users, multicast-style traffic replication and the information-aware placement model force both formulations to reuse the same high-utility paths; consequently, balancing the load no longer carries a significant additional price tag.

\begin{figure}[tb]
    \centering
    \includegraphics[width=0.9\linewidth]{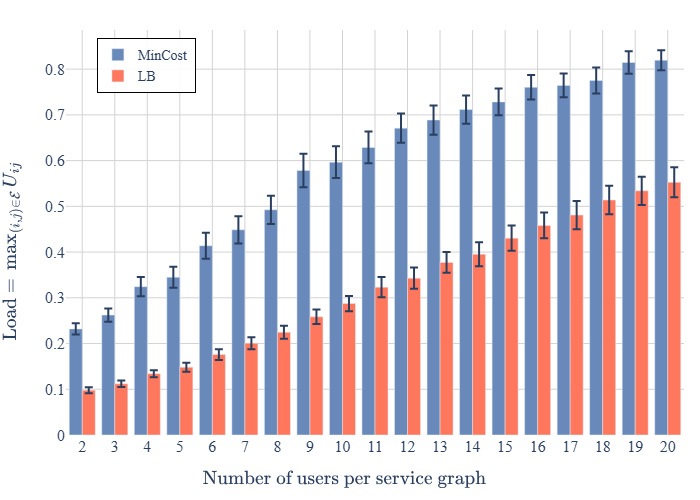}
    \caption{Peak link utilization versus number of users per service graph on an 8-node binomial cloud network.}
    \label{fig:cng8load}
        \vspace{-0.4cm}
\end{figure}

\begin{figure}[tb]
    \centering
    \includegraphics[width=0.9\linewidth]{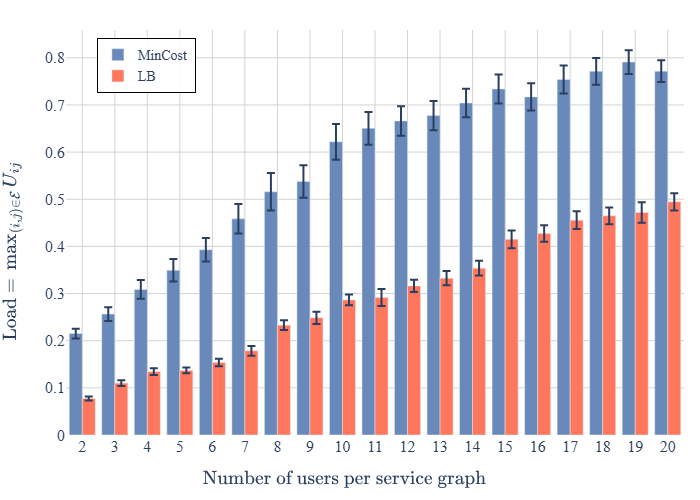}
    \caption{Peak link utilization versus number of users per service graph on a 16-node binomial cloud network.}
    \label{fig:cng16load}
        \vspace{-0.4cm}
\end{figure}

We now focus on load balancing. Figures ~\ref{fig:cng8load} and \ref{fig:cng16load} report the maximum link-utilization ratio achieved by the two MILP formulations on the 8- and 16-node CNGs, respectively. In both settings, and as expected, \emph{LB} keeps the maximum-utilized link markedly below that of \emph{MinCost}. Specifically, with 2 to 7 users, \emph{LB} achieves a load between 0.16 and 0.2, whereas the \emph{MinCost} achieves a cost between 0.2 and 0.45 (25\% to 55\% more). We also observe that these percentages increase significantly up to 63\% (0.47-0.5 for \emph{LB} against 0.76-0.79 for \emph{MinCost}) when the number of users in the SG is between 18-20 users. This demonstrates that while \emph{LB} may incur an additional cost of roughly 700 units per user on average, that cost can be compensated by realizing a more balanced load across the network. 

In summary, the analysis highlights a clear trade-off between cost efficiency and load balancing. While the \emph{MinCost} formulation yields lower per-user costs, especially at smaller scales, the \emph{LB} formulation achieves significantly better load distribution across the network. As the number of users increases, the gap in cost narrows, making \emph{LB} a compelling choice in scenarios where network stability and QoE are critical.


\section{Conclusion}\label{sec:conclusion}
In this paper, we study the end-to-end service provisioning for time-critical MM sessions over 5G/6G infrastructures. Starting from a detailed analysis of interactive virtual-concert workflows, we $(i)$ distill their functional dependencies into a service graph (SG), $(ii)$ formalize the MM processor placement and routing problem as a CNFlow  problem, $(iii)$ present MuMeNet, a discrete-event simulator that reproduces sub-millisecond cross-stream synchronization under realistic network impairments, and $(iv)$ formulate a Mixed-Integer Linear Program (MILP) that embeds MM SGs onto arbitrary CNGs under two different objectives: Minimizing overall cost, and minimizing maximum resource utilization. On the experimental side, integrating MuMeNet with commercial cloud/edge platforms will allow empirical validation of the simulator assumptions and extend the impairment models. In summary, this study lays the theoretical and methodological groundwork for delivering latency-critical, multisensory musical interactions at Metaverse scale. We believe that MuMeNet will accelerate reproducible research on ultra-low-latency multimedia systems, bridging the methodological gap between networking, edge computing, and musical-interaction communities.

\section*{Acknowledgment}
This work is supported by MUSMET project funded by the EIC Pathfinder Open scheme of the European Commission (grant agreement n. 101184379), and has received funding from the
Swiss State Secretariat for Education, Research and Innovation (SERI).

\bibliographystyle{IEEEtran} 
\bibliography{references}

\end{document}